\documentclass[twocolumn,preprintnumbers,aps,pra,amsmath,amssymb,floatfix,superscriptaddress]{revtex4}  

\usepackage{graphicx}
\usepackage{dcolumn}
\usepackage{bm}
\usepackage{amssymb}

\begin{document}

\title{Generation of Arbitrary Frequency Chirps with a Fiber-Based Phase Modulator and Self-Injection-Locked Diode Laser}

\author{C.E. Rogers III, M.J. Wright,\footnote{Present address: Institut f\"ur Experimentalphysik, Universit\"at Innsbruck, Technikerstra$\ss$e 25, 6020 Innsbruck, Austria} J.L. Carini, J.A. Pechkis, and P.L. Gould}

\email{phillip.gould@uconn.edu}

\affiliation{Department of Physics, University of Connecticut, Storrs, CT 06269, USA}

\date{\today}

\begin{abstract}

   We present a novel technique for producing pulses of laser light whose frequency is arbitrarily chirped. The output from a diode laser is sent through a fiber-optical delay line containing a fiber-based electro-optical phase modulator. Upon emerging from the fiber, the phase-modulated pulse is used to injection-lock the laser and the process is repeated. Large phase modulations are realized by multiple passes through the loop while the high optical power is maintained by self-injection-locking after each pass. Arbitrary chirps are produced by driving the modulator with an arbitrary waveform generator.
   
\end{abstract}


\maketitle 
\section{Introduction}
   In a variety of applications, it is desirable to be able to exert rapid and arbitrary control over the frequency of a laser, while minimizing the associated variations in intensity. Diode lasers, both free-running and in external cavities, are particularly amenable to rapid frequency control via their injection current.\cite{Watts 86} However, current modulation produces not only frequency modulation, but also intensity modulation, which is often undesirable. This issue has been addressed by injection-locking a separate laser with the modulated light.\cite{Wright 04} The frequency modulation is faithfully followed, while the intensity modulation is suppressed. Linear chirps up to 15 GHz/$\mu$s,\cite{Wright 04} as well as significantly boosted output powers \cite{Wright 04, Repasky 01} have been achieved in this manner. Other techniques for rapid tuning include electro-optical crystals located within the diode laser's external cavity \cite{Boggs 98, Menager 00, Repasky 01, Levin 02} and fiber-coupled integrated-optics waveguides to phase-modulate the laser output.\cite{Troger 99, Reibel 02B, Thevenaz 04} In the present work, we combine two key elements to produce pulses of arbitrarily frequency-chirped light: 1) an electro-optical waveguide phase modulator, located in a fiber loop and driven with an arbitrary waveform generator; and 2) optical self-injection-locking after each pass through the loop.\cite{Troger 99, Thevenaz 04} Multiple passes around the loop allow large changes in phase to be accumulated, and the self-injection-locking maintains a high output power. We expect such controlled light to be useful in applications such as adiabatic population transfer,\cite{Liendenbaum 89} coherent transients,\cite{Pietilainen 98, Li 99} atom optics,\cite{Bakos 06, Miao 06} ultracold collisions,\cite{Wright 05} radio-frequency spectrum analyzers based on spectral hole burning,\cite{Menager 01} optical coherent transient programming and processing,\cite{Merkel 98} and high-bandwidth spatial-spectral holography.\cite{Reibel 02A}
   \\
   \\
   
\section{Experimental Set-up}

   A schematic of the experimental set-up is shown in Fig. 1.  The central laser is a free-running diode laser (FRDL), Hitachi HL7852G, with a nominal output of 50 mW at 785 nm. Its temperature is reduced and stabilized in order to provide a wavelength near the Rb D$_{2}$ line at 780 nm. The output of this laser is sent through an optical isolator in order to prevent undesired optical feedback, and then through two acousto-optical modulators (AOM2 and AOM3) whose purpose will be discussed below. The beam is then coupled into a 40 m long single-mode polarization-maintaining fiber delay line to prevent overlap of successive pulses as they propagate around the loop. Within the loop is a fiber-coupled integrated-optics phase modulator. This device, EOSpace model PM-0K1-00-PFA-PFA-790-S, is a lithium niobate waveguide device capable of modulation rates up to 40 Gbit/s when properly terminated with 50 $\Omega$. Our device is unterminated in order to allow higher voltages (e.g., we use up to 10 V) to be applied. Upon emerging from this fiber loop, the light is coupled back into the FRDL in order to (re) injection-lock it.
   
\begin{figure}
\centerline{\includegraphics{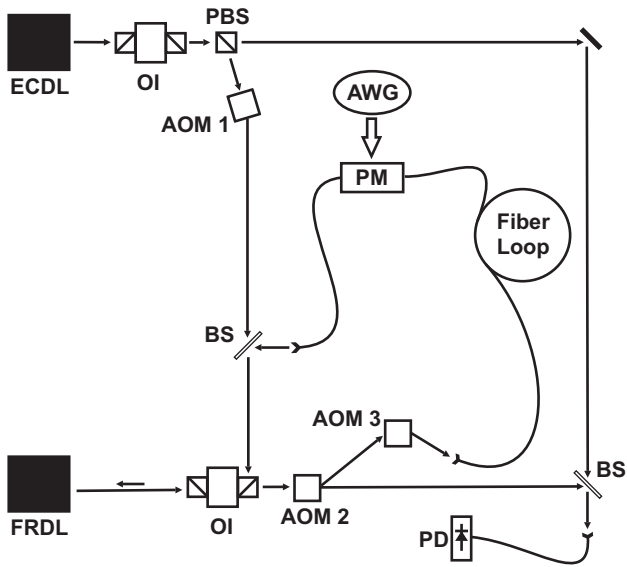}}
\caption{\label{fig:epsart} Schematic of the apparatus. The free-running diode laser (FRDL) is initially injection-locked by a seed pulse originating from the external-cavity diode laser (ECDL) and switched on by acousto-optical modulator AOM1. The injection-locked output pulse from the FRDL is switched into the fiber loop by AOM2. The frequency shift produced by AOM2 is compensated for by AOM3. The fiber loop is connected to a phase modulator (PM) driven by an arbritrary waveform generator (AWG). The phase-modulated pulse (re) injection locks the FRDL and the loop cycle is repeated. After N passes through the loop, the pulse is combined with the ECDL output on a fast photodiode (PD) for heterodyne analysis. Beamsplitters (BS), polarizing beamsplitters (PBS), and optical isolators (OI) are also shown.}
\end{figure}
   
   To initialize the FRDL frequency, we use a seed pulse from a separate external-cavity diode laser \cite{Ricci 95} (ECDL) to injection-lock it. The ECDL can have its frequency stabilized to a Rb atomic resonance using saturated absorption spectroscopy. The seed pulse, typically 170 ns in width, is generated using AOM1. After this initial seeding, the ECDL is completely blocked and the FRDL then self-injection-locks with the pulses of light emerging from the fiber loop. The two injection sources, which are not present simultaneously, are merged on a beamsplitter. This combined beam is directed into the FRDL through one port of the output polarizing beamsplitter cube of its optical isolator, thereby insuring unidirectional injection. Injection powers of 250 $\mu$W are typically used.

The timing diagram for the pulse generation is shown in Fig. 2. The FRDL emits light continuously, but injection-locks to the ECDL only during the brief seed pulse. AOM2 is pulsed on in order to switch this pulse into the fiber loop. This first-order beam is frequency shifted by the 80 MHz frequency driving AOM2. AOM3, which is on continuously, provides a compensating frequency shift. Without this compensation, a large frequency change would accumulate after multiple passes around the loop. Such a controllable frequency offset may be desirable for some applications. After passing through the fiber, the pulse enters the phase modulator. The desired modulation is imprinted on the pulse with an 80 MHz (200 MSa/s) arbitrary waveform generator (AWG): Agilent 33250A. The pulse then exits the fiber and (re) injection-locks the FRDL. The resulting pulse emerging from the FRDL is an amplified version of the phase-modulated pulse. It is sent through the loop again, in exactly the same manner as the original pulse, for further phase modulation. The switching of AOM2 and the voltage provided to the phase modulator by the AWG are synchronized to the 221 ns cycle time of the entire loop using a pulse/delay generator. This ensures that phase changes for each pass accumulate optimally. After the desired number of cycles through the loop, AOM2 is switched off, opening the loop and sending the pulse to the diagnostics and/or experiment. The entire sequence can be repeated at a rate determined by the loop time and the number of passes around the loop.

\begin{figure}
\centerline{\includegraphics{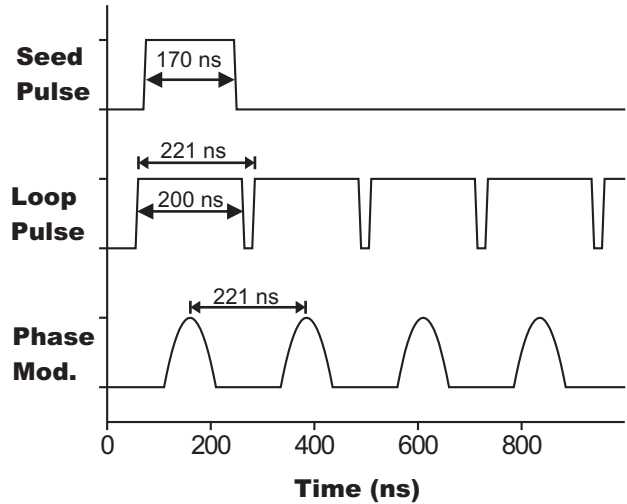}}
\caption{\label{fig:epsart} Timing diagram for chirped pulse generation. The seed pulse, generated by AOM1, initiates the process. Subsequent pulses of light from the FRDL, generated by AOM2, represent the multiple passes through the fiber loop. The desired phase modulation is applied synchronously during each pass.}
\end{figure}

Our main diagnostic is to combine the frequency-chirped pulse with the fixed-frequency light from the ECDL and measure the resulting heterodyne signal with a fiber-coupled fast photodiode and 500 MHz oscilloscope. We note that a fourth AOM outside the loop (not shown in Fig. 1) would allow the desired portion of the final chirped pulse to be selected and sent to the experiment. Because the initial seed pulse is typically shorter than the pulses propagating around the loop (see Fig. 2), there are portions of the output pulse during which the FRDL is not injection-locked at the desired frequency. We intentionally set the unlocked FRDL frequency far enough from that of the ECDL to ensure that only the desired portions of the pulse are visible in the heterodyne signal. Offsets ranging from 3 GHz to 600 GHz have been utilized, with smaller offsets providing more robust injection locking. For applications where light far from the ECDL frequency has no adverse effects, selection by the fourth AOM may not be necessary. 

An important advantage of our scheme is the fact that the injection locking amplifies the pulse to the original power level after each cycle, thereby allowing an arbitrary number of passes (we have used more than 20) through the modulator. This amplification is also important because the time-averaged optical power seen by the modulator must be limited (e.g., to $<$5 mW at our operating wavelength) to avoid photorefractive damage. We require only enough power in the fiber output, typically 750 $\mu$W, to robustly injection-lock the FRDL after each pass.

\section{Results}

To verify the fidelity of the injection locking, we perform the following test. With no voltage applied to the phase modulator, we pass a pulse through the loop 15 times before examining its heterodyne signal. Since the initial seed pulse is shifted 80 MHz by AOM1, and the shifts from AOM2 and AOM3 are set to cancel, we expect that the beat signal will be sinusoidal at 80 MHz. This is indeed the case, as shown in Fig. 3.

\begin{figure}
\centerline{\includegraphics{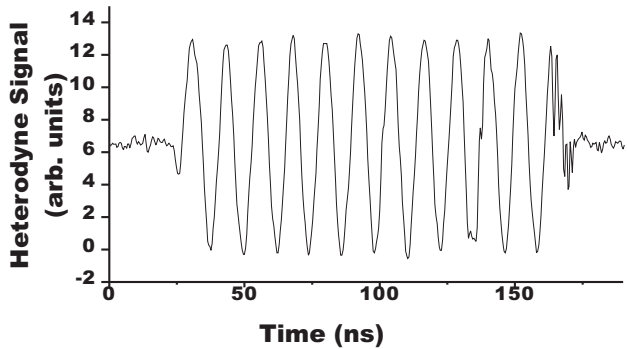}}
\caption{\label{fig:epsart} Heterodyne signal between the ECDL and the injection-locked FRDL pulse after 15 passes through the loop. No phase modulation is applied, so the 80 MHz beat signal is due to the frequency shift of AOM1. }
\end{figure}

The time varying frequency f(t) of a pulse is related to the modulated phase $\varphi$(t) by: 
\begin{equation}
\setcounter{equation}{1}
f(t)=f_{0}+(1/2\pi)(d\varphi/dt)	
\end{equation}
where f$_{0}$ is the original carrier frequency (in Hz). The phase change produced by N passes through the modulator is linear in the applied voltage V with a proportionality constant characterized by V$_{\pi}$:
\begin{equation}
\setcounter{equation}{2}
\Delta\varphi=N\pi(V/V_{\pi}).	
\end{equation}
We measure V$_{\pi}$ by applying a linear voltage ramp of 8 V in 100 ns and measuring the resulting frequency shift of 280 MHz after N=10 passes through the loop, as shown in Fig. 4. This yields V$_{\pi}$ = 1.4 V, somewhat more efficient than the specified value of 1.8 V.

\begin{figure}
\centerline{\includegraphics{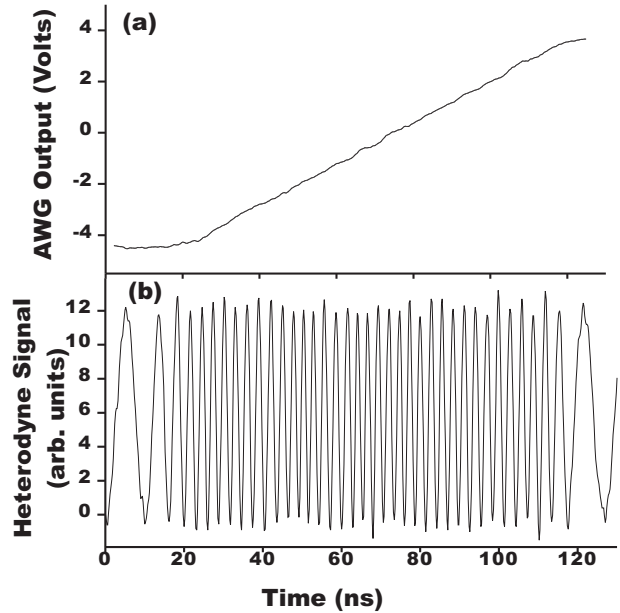}}
\caption{\label{fig:epsart} (a) Linearly varying output of the AWG which drives the phase modulator. (b) Heterodyne signal between the ECDL and the injection-locked FRDL pulse after 10 passes through the loop. The 360 MHz beat signal reflects the 80 MHz frequency shift of AOM1 as well as that due to the linear phase modulation.}
\end{figure}
	
	In order to produce a linear chirp, the phase change should be quadratic in time, requiring a quadratic voltage: V(t) = $\alpha$t$^{2}$. A series of increasing and decreasing quadratics, matched at the boundaries, is programmed into the AWG. This output voltage, together with the heterodyne signal and the resulting frequency as a function of time, are shown in Fig. 5. The inverse of the local period of the heterodyne signal, determined from successive minima and maxima, is used as the measure of frequency. Linear fits to the decreasing and increasing frequency regions yield chirp rates of -36 and +37 GHz/$\mu$s, respectively. These match well to the value of 38 GHz/$\mu$s expected from the programmed waveform and the value of V$_{\pi}$. We note that the chirp shown here is achievable only with multiple passes due to the input voltage limits of the modulator. However, if a given chirp range $\Delta$f is to be achieved in a time interval $\Delta$t, the required voltage change, $\Delta$V = $\alpha$($\Delta$t)$^{2}$ = (V$_{\pi}$/N)($\Delta$f$\Delta$t), is proportional to $\Delta$t, indicating that faster chirps are easier to produce.
	
\begin{figure}
\centerline{\includegraphics{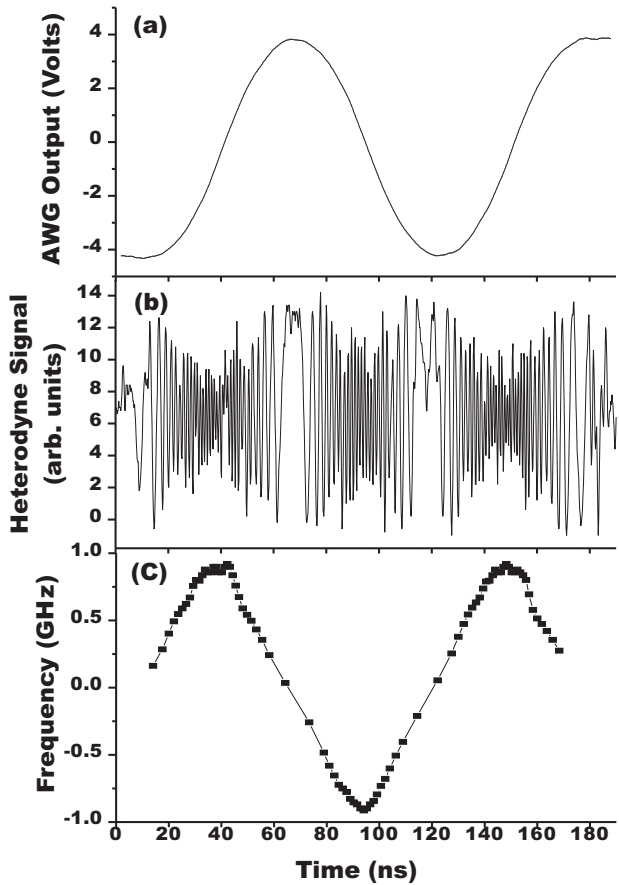}}
\caption{\label{fig:epsart} (a) Quadratically varying (alternately positive and negative) output of the AWG. (b) Heterodyne signal between the ECDL and the injection-locked FRDL pulse after 10 passes through the loop. (c) Frequency vs. time derived from (b).}
\end{figure}

	As an example of an arbitrary chirp, we show in Fig. 6 the result of a phase which varies quadratically in time with a superimposed sinusoidal modulation. The resulting frequency as a function of time, shown in (d), has the expected linear plus sinusoidal variation and matches quite well the numerical derivative of the AWG output, shown in (b). Although we have not yet explored this avenue, it should be possible to correct for imperfections in the AWG and/or the response of the phase modulator by measuring the chirp and adjusting the programmed waveform to compensate.     
\\
\\
\\	
\\
\\
\\

\begin{figure}
\centerline{\includegraphics{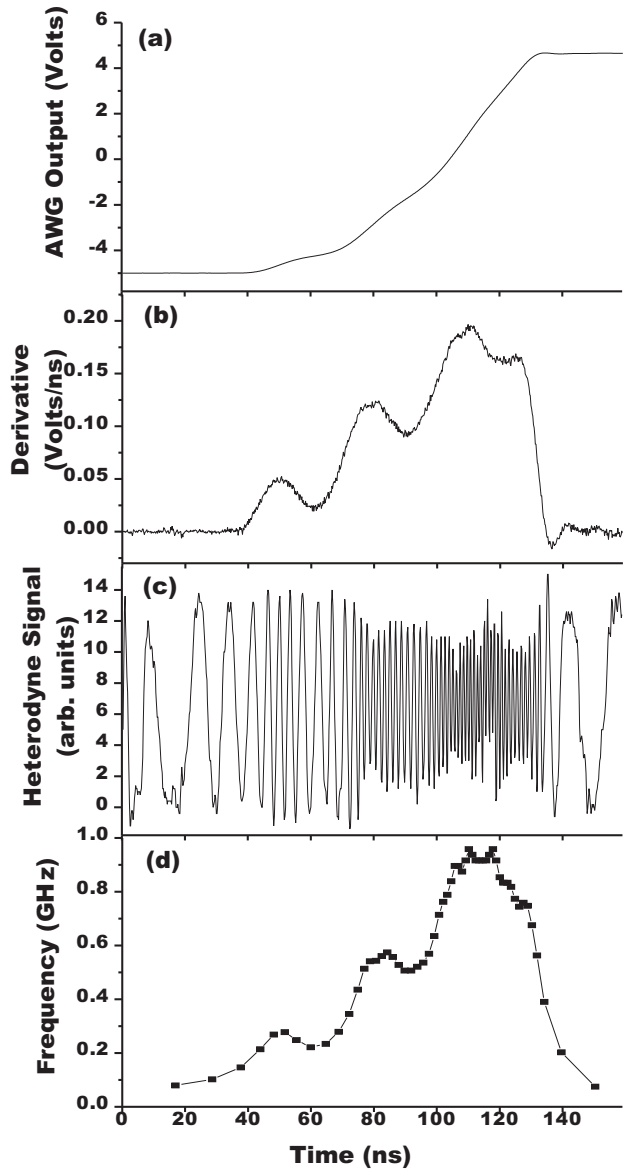}}
\caption{\label{fig:epsart} (a) Quadratic plus sinusoidal output of the AWG. (b) Numerical derivative of the AWG output. (c) Heterodyne signal between the ECDL and the injection-locked FRDL pulse after 13 passes through the loop. (d) Frequency vs. time derived from (c). Note the close correspondence between (b) and (d).}
\end{figure}

\section{Conclusion}

In summary, we have described a novel technique for producing pulses of light with arbitrary frequency chirps. The method is based on multiple passes through a fiber-based integrated-optics phase modulator driven by an arbitrary waveform generator, with self-injection locking after each pass. Our work has utilized light at 780 nm, but the chirping concept should work for a variety of wavelengths. We have shown examples of frequency shifts, linear chirps, and linear plus sinusoidal frequency modulations. We have yet to explore the limitations of this scheme. We are presently limited in modulation speed by the waveform generator, and our heterodyne diagnostic is limited by the bandwidths of both the oscilloscope (500 MHz) and the photodiode (1 GHz). For faster modulations, synchronization of successive passes will become more critical, but this can be adjusted either electronically or by the optical path length. We note that the phase modulation need not be identical for each pass, adding flexibility to the technique. At some point, the injection locking will not be able to follow the modulated frequency, but we see no evidence of this at the linear chirp rates of $\sim$40 GHz/$\mu$s (and corresponding chirp range of $\sim$2 GHz) which we have so far achieved. 

It is interesting to compare our scheme with pulse shaping in the femtosecond domain.\cite{Weiner 00} With ultrafast pulses, there is sufficient bandwidth to disperse the light and separately adjust the phase and amplitude of the various frequency components (e.g., with a spatial light modulator) before reassembling the shaped pulse. Our time scales are obviously much longer (e.g., 10 ns - 100 ns), and we control the phase directly in the time domain. A logical extension of our work would be to independently control the amplitude envelope with a single pass through a fiber-based integrated-optical intensity modulator. As with femtosecond pulse shaping and its application to coherent control, time-domain manipulations of phase and amplitude should be amenable to optimization via genetic algorithms.

\section{Acknowledgements}

This work was supported in part by the Chemical Sciences, Geosciences and Biosciences Division, Office of Basic Energy Sciences, U.S. Department of Energy. We thank Niloy Dutta for useful discussions and EOSpace for technical advice regarding the phase modulator.

\end{document}